\documentclass{article}
\usepackage{latexsym}
\newcommand{\ba}{\begin{eqnarray}}
\newcommand{\ea}{\end{eqnarray}}
\newcommand{\nn}{\nonumber\\}
\begin{document}
\title {The $e^+e^-$ pair production in relativistic ions collision  and its
correspondence to electron scattering}
\author{ A.N. Sissakian,A.V. Tarasov, H.T.Torosyan\footnote{On leave of absence from
Yerevan Physics Institute}, O.O.Voskresenskaya}
\maketitle
\begin{center}
Joint Institute for Nuclear Research, 141980 Dubna, Russia
\end{center}
\begin{center}
\bf {Abstract}
\end{center}

It is shown that the processes of electron(positron) scattering
and $e^+e^-$ pair production in the Coulomb fields of two
relativistic ions, can be expressed through the amplitudes of
lepton scattering off the  ion Coulomb field via  Watson
expansion. We obtained the compact expressions for these
amplitudes valid in the high energy limit and discuss the crossing
symmetry relations among the considered processes.

\section{Introduction}
A lot of work has been done in past years ~[1-12] in
 the investigation  of lepton pair production
in the Coulomb fields of two colliding relativistic ions with
charge numbers $Z_1,Z_2$ \ba
 Z_1+Z_2\to e^+e^-+Z_1+Z_2
\ea The main goal of this investigations is the attempt to obtain
the compact expression for the amplitude of process (1) accounting
for final state interaction of produced pair with ions in all
orders of fine structure constant
$\alpha=\frac{e^2}{4\pi}$.\\
The solving of this issue can help to understand very important
and unsolved problem of accounting the final state interaction of
quarks and gluons in QCD. Unfortunately even in QED up to now this
problem is not solved due to its complexity and so any progress in
this direction is very useful.\\
The investigation of the process (1) becomes much more simple in
the ultrarelativistic limit because of strong Lorentz contraction
of electromagnetic fields of ions moving with the velocity  close
to the speed of light. An example of such simple solution which
leads to an amplitude  differing from the Born one  only by phase
factor was obtained  in ~\cite{SW,BM,ERSG} . By virtue of
procedure of the crossing symmetry the amplitude of  the process
(1) have been obtained from so-called ``exact'' result for
amplitude of electron scattering in the Coulomb field of two
colliding ions, which has been extracted from ``exact'' solution
of Dirac equation
for electron in this field.\\
The further analysis of this problem in the framework of more
familiar Feynman  diagrams  leads to the conclusion that the
result of ~\cite{SW,BM,ERSG} is incorrect.This leads some authors
~\cite{ERG,BGK} to surprising conclusion that crossing symmetry
property is violated  beyond the Born approximation.\\
Despite the essential progress achieved in this direction
~\cite{BGKN1,BGKN2}, the authors of which succeed in the summing
some class of main diagrams, the general structure of amplitudes
for electron scattering in the Coulomb field of two relativistic
nuclei and $e^+e^-$ production in this fields has not been
established yet even in the limit of ultrarelativistic energies.
Taking into account the importance of this problem  and growing
interest to this issue from the scientific community we would like
to do some remarks, which as we hope will be useful  for
understanding of this problem.
\section {Electron scattering and pair production in the Coulomb
fields of two colliding nucleus }
 The amplitudes of electron scattering and lepton pair production
 in the arbitrary external electromagnetic field $A_{\mu}(x)$ can be
cast in the following form
\ba
 A^{(scat)}&=& \bar u(p_f)\int d^4 x _1 d^4 x_2 e^{ip_i x_1  - ip_f x_2 }
 T(x_2,x_1)u(p_i)\nn
 A^{(prod)}&=& \bar u(p_2)\int d^4 x_1 d^4 x_2 e^{ - ip_1x_1 -
ip_2x_2 } T(x_2 ,x_1)v(p_1)
 \ea
 where the function $ T(x_2 ,x_1) $
 is the same in both cases and obeys  the following equation
\ba
  T(x_2,x_1) = V(x_2,x_1) - \int d^4 x d^4 x'V(x_2 ,x)G(x,x')T(x',x_1)
   \ea
  or in the short notation
 \ba
T& =& V - V \otimes G \otimes T \nn
 V\left( {x_2 ,x_1 } \right)& =& e\gamma _\mu
{\rm A}_\mu  \left( {x_1 } \right)\delta ^{\left( 4 \right)}
\left( {x_2  - x_1 } \right)\nn G\left( {x,x'} \right)& =&
\frac{1}{{\left( {2\pi } \right)^4 }}\int {d^4 k\frac{{\hat k +
m}}{{k^2  - m^2  + i0}}} e^{ - ik\left( {x - x'} \right)} \ea
where  $\gamma_{\mu}$ and $u(p),v(p)$ are Dirac matrixes  and
spinors.\\
 From this relations it follow that for the two center
problem ($ {\rm A}_\mu \left( x \right) = {\rm A}_\mu ^{(1)}
\left( x \right) + {\rm A}_\mu ^{(2)} \left( x \right) $) the
amplitude $ T(x_2,x_1)$ can be represented in the form of
infinitive Watson series ~\cite{W} \ba T &=& T^{(1)} + T^{(2)} -
T^{(1)}  \otimes G \otimes T^{(2)}  - T^{(2)} \otimes G \otimes
T^{(1)}  \nn &+& T^{(1)} \otimes G \otimes T^{(2)} \otimes G
\otimes T^{(1)}+ T^{(2)} \otimes G \otimes T^{(1)} \otimes G
\otimes T^{(2)} ... \ea
 where $ T^{(1,2)}$ obey the equations
\ba
 T^{(1)}&=& V^{(1)}  - V^{(1)}  \otimes G \otimes T^{(1)}\nn
  T^{(2)}&=&V^{(2)}  - V^{(2)}  \otimes G \otimes T^{(2)}
\ea
In  high energy limit when Lorentz factor of colliding ions $
\gamma=\frac{E}{M} \to \infty$  this equations can be solved with
the result
 \ba
 T^{(1)}&(x_2,x_1)=\gamma _+[U_1(x_1)\delta^4(x_2 - x_1)
 + \frac{i}{2\pi}\delta^2(\vec x_2 -\vec x_1)U_1(x_2)U_1(x_1)
 \exp{\left(i\int\limits_{x_{1+}}^{x_{2+}}U_1(x)dx_+\right)}\nn
&\times\int\limits_{-\infty }^{\infty}dk_+\left(\theta (k_+)\theta
(x_{2+}-x_{1+})-\theta (-k_+)\theta (x_{1+} -x_{2+})\right)
 \exp \left(-\frac{ik_+(x_{2-}-x_{1-})}{2}\right)]
\ea
\ba
 T^{(2)}&(x_2,x_1)=\gamma _-[U_2(x_1)\delta^4(x_2 - x_1)
 + \frac{i}{2\pi}\delta^2(\vec x_2 -\vec x_1)U_2(x_2)U_2(x_1)
 \exp{\left(i\int\limits_{x_{1-}}^{x_{2-}}U_2(x)dx_-\right)}\nn
&\times\int\limits_{-\infty }^{\infty}dk_-\left(\theta (k_-)\theta
(x_{2-}-x_{1-})-\theta (-k_-)\theta (x_{1-} -x_{2-})\right)
 \exp \left(-\frac{ik_-(x_{2+}-x_{1+})}{2}\right)]
 \ea
where
  \ba
 U_1(x)& =& e\gamma \Phi _1 \left({\sqrt {\left(\vec b_1- \vec x\right)^2 + \gamma ^2 x_ + ^2 } }
\right) \nn
 U_2(x)&=& e\gamma \Phi _2 \left(
{\sqrt {\left(\vec b_2-\vec x\right)^2 + \gamma ^2 x_ - ^2 } }
\right) \ea
 Here $\Phi_{1,2}(r)$ are the Coulomb
potentials of ions $Z_{1,2}$ ;$\gamma_{\pm} =\gamma_0 \pm
\gamma_z$ is the Dirac matrixes  and we use the lightcone
definition of momenta and coordinates  $k_{\pm}=k_0 \pm
k_z,x_{i\pm}=x_{i0}\pm x_{iz}$.~$\vec b_1,\vec b_2$ are the the
impact parameters of ions and $\vec x_1,\vec x_2$ the transverse
coordinates of relevant four-vectors.There are no other
simplifications in high energy limit. In particular there are no
truncation of infinite Watson series in contrary to the statement
done in ~\cite{ERG}.
\section{The crossing symmetry relations}
Let us discuss the property of crossing symmetry using as example
the simplest crossed reactions - electron and positron scattering
in the Coulomb field of ion.
 The amplitudes of these reactions read
  \ba
 A(e^-Z\to e^-Z)&=& \bar u(p_f^{-})T^{(1)}(p_f^{-},p_i^{-})u(p_i^{-})\nn
 A(e^+Z\to e^+Z)&=&-\bar v(p_i^{+})T^{(1)}(-p_i^{+},-p_f^{+})v(p_f^{+})\nn
 \ea
 where
\ba
 T^{(1)}(p_2,p_1)&=&\int d^4 x_1 d^4 x_2 e^{ip_1 x_1  - ip_2 x_2 }
T(x_2 ,x_1 )=(2\pi )^2 \delta (p_{2 + } - p_{1 + })\nn
&\times&\gamma_+[\theta(p_{1 + })f^+(\vec p_2 -\vec p_1) -
\theta(- p_{1+})f^-(\vec p_2 - \vec p_1)]\\
 f^{\pm}(\vec q)&=&\frac{i}{2\pi}\int d^2x e^{i\vec q  \vec x }
 [1-e^{\pm i\chi(\vec b - \vec x)}]\nn
 \chi (\vec b-\vec x)&=&e\int\limits_{ - \infty }^\infty\Phi
 \left (\sqrt{( \vec b  - \vec x )^2 + z^2}\right ) dz\nonumber
 \ea
 The crossing symmetry for reactions (10) means that the
amplitudes for electron and positron scattering in the Coulomb
field  are expressed through universal  function $T(p_2,p_1)$
calculated at different values of its arguments. If this function
would be analytical function of its arguments then it would be
possible to express one amplitude through another using a simple
substitution $p_i^{-} \to - p_f^{+};p_f^{-}\to - p_i^{+}$ and
trivial changing of the spinors $ u\to v $. But discontinuity of
$T(p_2,p_1)$ in the variable $p_+$  doesn't allow to do this in
general case.Such substitution can be done only on the level of
Born approximation as is easily seen from (11).\\
For the same reason the amplitude of  pair production cannot be
derived from the amplitude of electron scattering  by procedure of
crossing symmetry beyond the Born approximation.
\section{Conclusions}
There are two wrong points in the derivation of  expressions for
$e^+e^-$ pair production amplitude in ~\cite{SW,BM,ERSG}. The
first one is the oversimplified expression for electron scattering
amplitude.Really the authors omit all higher terms of Watson
expression (5) except the four first ones. The other mistake which
leads the authors  to  wrong statement on absence of Coulomb
corrections to the Born approximation is  the wrong application
of crossing symmetry property as we explained above. \\
Thus the  problem of  final state interaction to all orders in the
fine structure constant even in abelian theory is a complex issue
and demands the further and deeper investigation  which  will be
done elsewhere. The authors gratefully acknowledge fruitful
discussion with S.R.Gevorkyan, E.A.Kuraev and N.N.Nikolaev.

\end{document}